\documentstyle[prb,twocolumn,aps,epsfig]{revtex}

\newcommand {\cuni} {Cu$_{1-x}$Ni$_x$GeO$_3$ }
\newcommand {\cugeo} {C\lowercase {u}G\lowercase {e}O$_3$ }
\newcommand {\nilow} {N\lowercase {i}}

\title{Electron Spin Resonance of Ni-doped CuGeO$_3$ in the paramagnetic, 
spin-Peierls and antiferromagnetic states: Comparison with non-magnetic 
impurities}

\author{B. Grenier}
\address{D\'{e}partement de Recherche Fondamentale sur la Mati\`{e}re 
Condens\'{e}e, SPSMS-MDN, CEA/Grenoble, 17 rue des Martyrs, 38054 Grenoble, France\\
and Universit\'e Joseph Fourier Grenoble I, France}

\author{P. Monod}
\address{Laboratoire de Physique du Solide, Ecole Sup\'{e}rieure de Physique 
et Chimie, 10 rue Vauquelin, 75231 Paris, France}

\author{M. Hagiwara, M. Matsuda\cite{matsuda} and K. 
Katsumata\cite{katsumata} }
\address{RIKEN (The Institute of Physical and Chemical Research), Wako, 
Saitama 351-0198, Japan}

\author{S. Cl\'ement and J.-P. Renard}
\address{Institut d'Electronique Fondamentale, B$\hat{a}$t.220, 
Universit\'e Paris-Sud, 91405 Orsay, France}

\author{A.L. Barra}
\address{Laboratoire des Champs Magn\'etiques Intenses, CNRS/MPG, 24 rue 
des Martyrs, 38042 Grenoble, France}

\author{G. Dhalenne and A. Revcolevschi}
\address{Laboratoire de Physico-Chimie de l'Etat Solide, B$\hat{a}$t. 414, Universit\'e 
Paris-Sud, 91405 Orsay, France}

\date{\today}

\begin{document}
\twocolumn[\hsize\textwidth\columnwidth\hsize\csname @twocolumnfalse\endcsname

\maketitle
\begin{abstract}

We have performed Electron Spin Resonance measurements on single crystals 
of the doped spin-Peierls compounds CuGe$_{1-y}$Si$_y$O$_3$ and 
Cu$_{1-x}$M$_x$GeO$_3$ with M = Zn, Mg, Ni ($x, y\leq 0.1$). The first part 
of our experiments was performed in the paramagnetic and spin-Peierls 
phases at 9.5, 95 and 190 GHz. All non-magnetic impurities (Si, Zn and Mg) 
were found to hardly affect the position and linewidth of the single line 
resonance, in spite of the moment formation due to the broken chains. In 
contrast to Si, Zn and Mg-doping, the presence of Ni (S = 1) at low
concentration induces a spectacular shift towards high fields of the ESR
line (up to 40\% for x=0.002), together with a large broadening. This 
shift is strictly proportional to the ratio of Ni to Cu susceptibilities: 
Hence it is strongly enhanced below the spin-Peierls transition. We 
interpret this shift and the broadening as due to the exchange field induced by the Ni ions 
onto the strongly exchange coupled Cu spins. Second, the antiferromagnetic 
resonance was investigated in Ni-doped samples. The frequency vs magnetic 
field relation of the resonance is well explained by the classical theory 
with orthorhombic anisotropy, with $g$ values remarkably reduced, in 
accordance with the study of the spin-Peierls and paramagnetic phases. The 
easy, second-easy, and hard axes are found to be $a$, $c$, and $b$ axes, 
respectively. These results, which are dominated by the single ion 
anisotropy of Ni$^{2+}$, are discussed in comparison with those in the Zn- 
and Si-doped CuGeO$_3$.

\end{abstract}
\pacs{PACS: 75.10Jm, 76.30Da}
]


\section{Introduction}
\label{sec:I}

The quasi-one dimensional compound CuGeO$_3$ is the first inorganic system 
exhibiting a spin-Peierls (SP) transition\cite{Hase93a}. It has been 
extensively studied since large single crystals are available, which allows 
to perform various kinds of experiments. The magnetic structure of 
CuGeO$_3$ is as follows: the spin correlations are antiferromagnetic along 
the $b$ and $c$ axes and ferromagnetic along the $a$-axis. Below 
$T_{SP}\simeq 14.3$ K, the $S=\frac{1}{2}$ Heisenberg antiferromagnetic 
(AF) chains, running along the $c$-axis, become dimerized and an energy gap 
$\Delta = 2$ meV opens between the singlet ground state and the first 
excited triplet states. The SP transition is evidenced by a kink at 
$T_{SP}$ in the magnetic susceptibility and is clearly revealed by X-ray 
and neutron scattering\cite{Pouget94}.
A unique feature of this compound is that the effect of impurities can be 
studied by substituting impurity ions (Zn, Mg and Ni) for the Cu sites 
\cite{Hase93b,Hase95,Oseroff95,Lussier95,Coad96,Masuda98} and Si for the Ge 
sites \cite{Renard95}. These substitutions all induce a strong decrease of 
$T_{SP}$ and the appearance at lower temperature ($T < T_N < T_{SP}$) of a 
three-dimensional AF phase. The most intriguing feature in the impurity 
effects is the coexistence below $T_N$ of the SP state and the long-range 
AF ordering, which was extensively studied both 
experimentally\cite{Regnault95,Sasago96} and 
theoretically\cite{Fukuyama96}. Magnetization measurements performed in 
Cu$_{1-x}$M$_x$GeO$_3$ (with M = Zn, Mg, Ni) and CuGe$_{1-y}$Si$_y$O$_3$ 
single crystals have revealed the existence of a universal Temperature - 
Doping concentration phase diagram with a scaling factor 
$x=3y$\cite{Grenier98b}. The spin-Peierls temperature shows a linear 
decrease as the doping level increases, following the relation 
$T_{SP}(x,y)/T_{SP}(0) = 1 - 15 x = 1 - 44 y$, while the N\'eel temperature 
increases linearly from 0 at low doping levels up to a broad maximum around 
4.5 K for $x = 3 y \simeq 0.04$ and then gradually decreases. Thus, doping 
on the Ge site by Si impurities (spin 0) is three times more efficient than 
doping on the Cu site, either by magnetic (Ni: spin 1) or non-magnetic (Zn, 
Mg) impurities. For all non-magnetic impurities, the easy axis in the AF 
phase is the $c$-axis\cite{Lussier95,Renard95,Regnault95,Grenier98b,Hase96b,Hase96,Fronzes97,Nojiri97}. 
However, some deviations from this universal behavior are 
noted in the case of Ni concerning the AF phase. Indeed, the $T_N(x)$ curve 
for Ni is slightly lower\cite{Grenier98b} and the easy axis was found to be 
the $a$-axis from susceptibility\cite{Grenier98b,Koide96} and neutron 
scattering experiments\cite{Coad96}.

Electron spin resonance (ESR) studies of the paramagnetic and spin-Peierls 
phases have been performed by different authors in 
pure\cite{Oseroff94,Yamada96,Honda,Pilawa97}, Si-, and Zn-doped 
CuGeO$_3$\cite{Fronzes97,Hase98,Hassan98}, and the antiferromagnetic 
resonance has also been studied for these two 
substitutions\cite{Hase96,Fronzes97,Nojiri97}. But very little has been 
done using the ESR technique in Ni-doped CuGeO$_3$, besides the work of Glazkov et al.\cite{Glazkov98} where a large number of parameters is needed for the interpretation.

The first aim of this paper is to make a comparative ESR study of 
Cu$_{1-x}$M$_x$GeO$_3$ (with M = Zn, Mg, Ni) and CuGe$_{1-y}$Si$_y$O$_3$ 
compounds, to find out if the universal character of the 
Temperature-Concentration phase diagram holds also for ESR, especially 
when comparing non-magnetic and magnetic impurities. Extensive ESR 
measurements were performed in the paramagnetic and spin-Peierls phases on 
high quality single crystals, for various doping levels. The results 
obtained in the non-magnetic impurity doped compounds are briefly presented 
and those obtained in Ni-doped CuGeO$_3$ are focused on. For the latter, 
our results on the ESR line shift are analyzed in detail and a model is 
proposed to explain the observed effect.

The second aim of this paper is to study the antiferromagnetic resonance 
(AFMR) in Ni-doped CuGeO$_3$ samples, and to compare the results with those 
obtained in the case of non-magnetic impurities. ESR is indeed a crucial 
technique to study the magnetic anisotropy at a microscopic level. In both 
studies (ESR and AFMR), Ni substitution is shown to produce different 
effects on the spin resonance of CuGeO$_3$, as compared to the case of 
non-magnetic impurities.

\section{Experimental set up}
\label{sec:II}

The CuGe$_{1-y}$Si$_y$O$_3$ (y = 0, 0.002, 0.007, 0.0085, 0.015, 0.06, 
0.085) and Cu$_{1-x}$M$_x$GeO$_3$ (M = Zn with $x = 0.016$, 0.039, 0.01; M 
= Mg with $x = 0.01$, 0.03, 0.05; M = Ni with $x = 0.002$, 0.0085, 0.02, 
0.03, 0.04) single crystals used in this study were grown from the melt 
under oxygen atmosphere using a floating zone method associated with an 
image furnace. All samples were then analyzed using Inductively Coupled 
Plasma Atomic Emission Spectroscopy (ICP/AES). The doping levels that are 
reported here are thus the effective ones derived from the ICP/AES 
analysis, which are usually slightly lower than the nominal concentration.

The ESR measurements were performed between 4 and 300 K in all single 
crystals on a standard X-band spectrometer (Bruker ESP 300E), operating at 
a frequency $f \simeq 9.5$ GHz and in a field range $\pm 1.6$ T. With this 
apparatus, the measured signal is the field derivative of the absorption. 
The crystal of typical mass 5 to 10 mg can be mounted onto the sample 
holder in order that the magnetic field lies either in the $(a,b)$ or in 
the $(a,c)$ plane. Because of the huge broadening of the ESR line induced 
by Ni-doping, further measurements were performed on a home-made high field 
spectrometer. These experiments were crucial to interpret the observed 
phenomena. The measurements were performed between 4 and 250 K at 
frequencies 95 and 190 GHz, with the field parallel to the $c$-axis. The 
sample size was about 4$\times$8$\times$8 mm$^3$, corresponding to a mass 
of more than 500 mg. With this spectrometer, the measured ESR signal is the 
field derivative of a combination of the dispersion and absorption. The 
AFMR measurements were performed on two \cuni single crystals, with 
$x=0.03$ and 0.04, in the frequency ranges of 9, 20, 35, and 50 GHz and a 
temperature of 1.8 K ({\it i.e.} below $T_N$). The dimension of the 
crystals used in these experiments is about 3$\times$3$\times$2 mm$^3$. We 
used an X-band spectrometer (JOEL-JES-RE3X) for the experiments around 9 
GHz and a home-made spectrometer for higher frequencies. For the latter, 
microwaves are generated from two klystrons at 20 and 50 GHz and from a 
Gunn-oscillator at 35 GHz, the magnetic field ($B$) being produced by a 20 
T superconducting magnet from Oxford Instruments. The magnetic 
susceptibility measurements in the Ni-doped samples are also presented in 
this paper (for the other compounds, see Ref.\ \onlinecite{Grenier98b}): 
they were performed on a home-made SQUID magnetometer operating in the 
temperature range $1.8-300$ K and in the magnetic field range $0-8$ T.

\section{ESR in \cugeo doped with non-magnetic impurities}
\label{sec:III}

The temperature dependence of the magnetic susceptibility was first 
measured in all the samples whose ESR study is presented here (see Ref. 
\onlinecite{Grenier98a} for Si-doping and Ref. \onlinecite{Grenier98b} for Si, Zn and 
Mg-doping). We present briefly in this section the ESR results obtained in 
the case of non-magnetic impurities and discuss the linewidth, the shift, 
and the integrated intensity of the resonance line.

A wide series of CuGeO$_3$ samples doped with non-magnetic impurities Si, 
Zn, and Mg (see the list in sec.\ \ref{sec:II}) and a sample of pure 
CuGeO$_3$ were investigated. In all Si-, Zn- and Mg-doped CuGeO$_3$ 
compounds, a single lorentzian derivative ESR line was observed at X-band 
between 5 and 300 K along the three crystallographic axes, as for pure 
CuGeO$_3$. 

\vspace{0.2cm}
\begin{figure}
   \begin{center}
    \mbox{\epsfxsize=8cm \epsffile{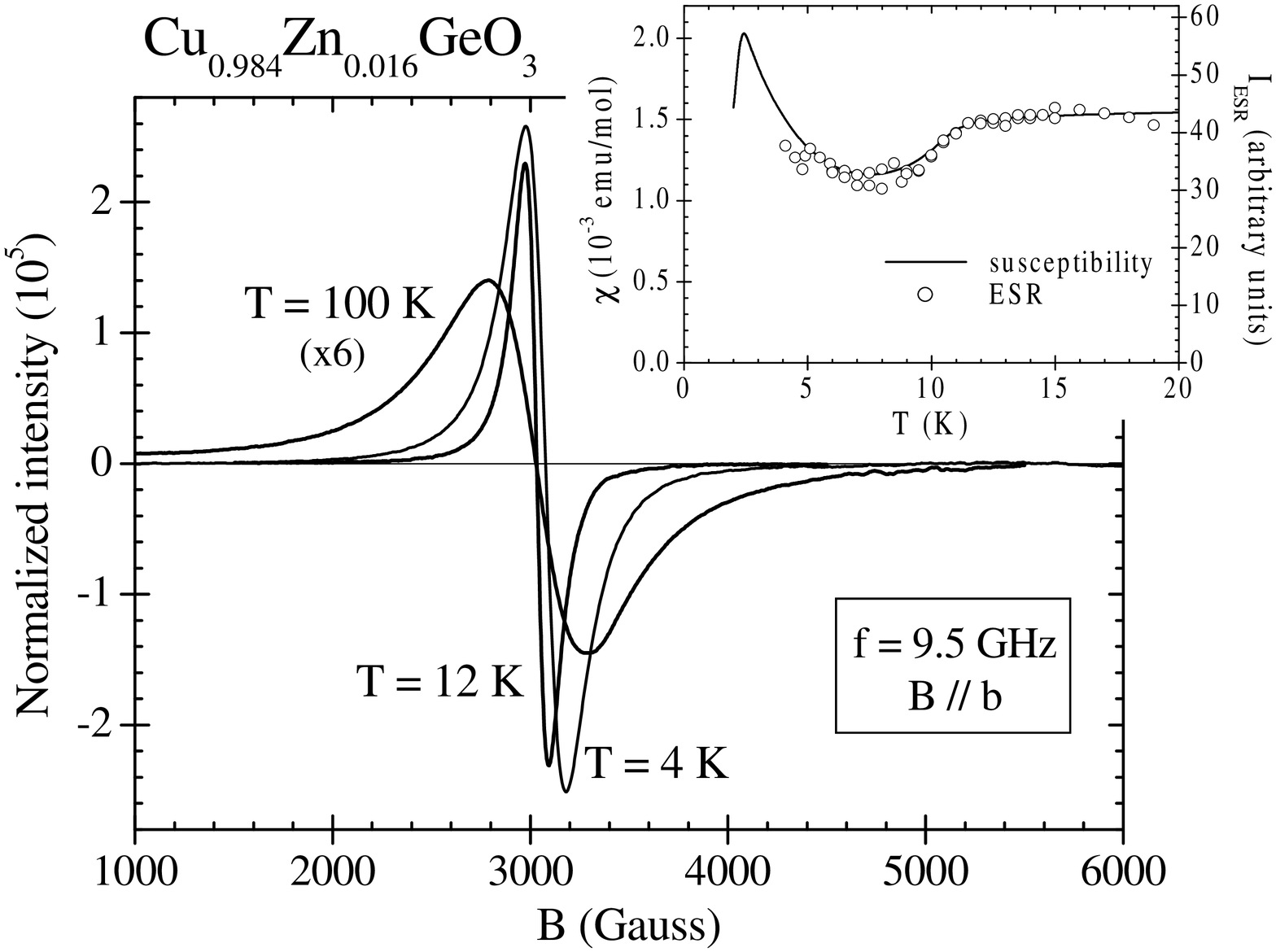}}
		\vspace{0.2cm}
    \caption{ESR spectra obtained at X-band, with $B$$\parallel$$b$, in 
Cu$_{0.984}$Zn$_{0.016}$GeO$_3$, above and below the SP temperature. The 
inset illustrates the temperature dependence of the (twice) integrated ESR 
intensity as compared to that of the static susceptibility.}
    \label{Fig1}
   \end{center}
\end{figure}

Figure\ \ref{Fig1} shows examples of spectra obtained in a 1.6\% 
Zn-doped sample for $B$$\parallel$$b$, {\it i.e.} for the field perpendicular 
to the chain axis, and at different temperatures, above and below 
$T_{SP}=10.4$ K. In all these samples and along the three crystallographic 
directions, the linewidth $\Delta B$ and its temperature dependence were 
found to be identical to those of pure CuGeO$_3$ above 15 
K\cite{Grenier99}, but not below: $\Delta B(T)$ depends on the N\'eel and 
SP temperatures, which vary with the doping level. As can be seen in Fig.\ 
\ref{Fig1}, a 1\% shift of the ESR line is observed only below $T_{SP}$. 
The same result was found in all CuGeO$_3$ samples doped with non-magnetic 
impurities and also in the pure compound: The $g$ values remain constant 
down to 10-15 K and become shifted at lower temperature by less than 2\%. 
The sign of the shift depends on the field orientation and on the 
substituent, which agrees well with the results of Hase {\it et al.} 
\cite{Hase98} obtained on pure, Zn- and Si-doped CuGeO$_3$. For pure and 
Si-doped CuGeO$_3$, the g-factor is found to increase along $b$ and to 
decrease along $a$ and $c$, as temperature decreases. For Zn- and Mg-doped 
CuGeO$_3$ the opposite behavior is observed for the three crystallographic 
directions.

The inset of Figure\ \ref{Fig1} shows the variation with temperature of the 
ESR (twice) integrated intensity $I_{ESR}$ as compared to that of the 
static susceptibility, in the 1.6\% Zn-doped CuGeO$_3$ sample. Since both 
curves can be superimposed, we infer that ESR at X-band cannot distinguish 
between triplet spins (spin-Peierls contribution) and spins freed by 
impurities (Curie-Weiss contribution)\cite{Grenier98b,Grenier98a}. We will 
come back to this point in Sec.\ \ref{sec:V.2}. A similar investigation carried out on 
Zn-doped CuGeO$_3$ (up to 5\% Zn) on powder samples at high field (between 
100 and 400 GHz) by Hassan et al.\cite{Hassan98} yields results 
qualitatively different from the previous investigations at lower frequency 
(see Ref. \onlinecite{Hase96,Fronzes97} and present work). In Ref.\onlinecite{Hassan98}, the appearance of two 
distinct powder pattern ESR spectra is interpreted as a separate 
contribution from the excited spin-Peierls triplet and the Zn-induced 
moments. Motivated by this apparent paradox, we have investigated the ESR in the
paramagnetic phase of a single crystal of CuGeO$_3$ : 5\% Zn at 95 GHz and 5 K.
Contrary to Ref.\onlinecite{Hassan98}, we observe for all field directions in the (a, b) plane
a single resonance line varying from $g_a = $2.145 to $g_b = $2.240, in
accordance with our X-band measurements. Thus, we are forced to conclude
that the additional ESR powder pattern observed by Hassan et al.\cite{Hassan98} is
somehow linked to the powdered nature of the investigated samples.

\section{Magnetic susceptibility in \nilow -doped \cugeo}
\label{sec:IV}

\begin{figure}
   \begin{center}
     \mbox{\epsfxsize=7.2 cm \epsffile{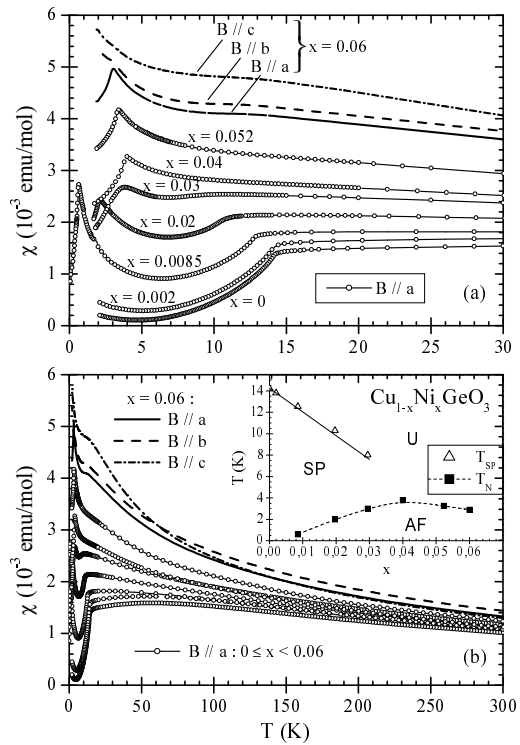}}
	\vspace{0.2cm}
   \caption{Temperature dependence of the magnetic susceptibility measured 
in \cuni samples in a 1 kGauss field applied along the $a$-axis (easy axis) 
for $0\leq x\leq 0.052$ and along the three crystallographic axes for 
$x=0.06$. The inset of figure (b) displays the $(T,x)$ phase diagram 
obtained from these measurements.}
   \label{Fig2}
   \end{center}
\end{figure}

Figure\ \ref{Fig2} shows the temperature dependence of the magnetic 
susceptibility measured in pure CuGeO$_3$ and in \cuni samples in the 
1.8-300 K temperature range (Fig.\ \ref{Fig2}b) and below 30 K (Fig.\ 
\ref{Fig2}a).

The data from Figure\ \ref{Fig2}a clearly show the decrease of the SP 
transition (which is visible up to $x=0.03$) and the occurrence of the AF 
transition at lower temperature, with $a$ being the easy axis. Indeed, the 
susceptibility measured with the external field along the $a$-axis shows a 
peak, around 3 K for $x=0.06$ (see solid line), while the susceptibilities 
measured with the external field along the $b$ and $c$ axes further increase below 3 K with 
decreasing temperature (see dash and dash-dot lines). The Ni$^{2+}$ ion, 
with $S=1$, is considered to participate to the long-range magnetic 
ordering since the magnitude of the Curie tail is one order of magnitude 
smaller than expected from 6\% Ni free spins, indicating that the 
interaction between the Cu$^{2+}$ and Ni$^{2+}$ ions is relatively large. 
The $(T,x)$ phase diagram shown in the inset of Fig.\ \ref{Fig2}b was 
obtained from these measurements. The values of the spin-Peierls and N\'eel 
temperatures (when they exist) are given in Table\ \ref{table1}. For 
$x=0.002$, the susceptibility measurement was performed only down to $T = 
1.8$ K, so that the N\'eel temperature (expected to be smaller than the 0.6 
K value obtained for $x=0.0085$) could not be determined.

Figure\ \ref{Fig2}b illustrates the large effect of Ni-doping on the global 
susceptiblity curve, due to the magnetic nature of the Ni$^{2+}$ ion. 
Indeed, the susceptibility increases more and more between 300 and 20 K as 
the Ni concentration increases. Interestingly, the susceptibilities along 
the $a$ and $b$ axes become smaller than the one along $c$ below $\simeq$ 
50 K. This anomalous temperature dependence was not observed in Si- and 
Zn-doped CuGeO$_3$ \cite{Renard95,Grenier98b,Hase95b}. This could originate 
from the single ion anisotropy of Ni$^{2+}$ ions $DS_z^2$ with $z$ along 
the direction of the distorted oxygen octahedron in the $(a,b)$ plane, as 
clearly observed with Mn$^{2+}$ (Ref. \onlinecite{GrenierMn}). This will be further 
discussed in the AFMR results presented in Sec. VI.

The temperature dependence of the susceptibility could be analyzed in the 
paramagnetic phase ($T>20$ K) by a crude model of Cu-Ni pairs, neglecting 
in a first step the single ion anisotropy of Ni$^{2+}$, yielding an 
estimate of the Ni-Cu antiferromagnetic exchange coupling of 25 K 
\cite{Grenier98b}. This value is to be compared to the strongest Cu-Cu 
coupling, $J_c = 120-180$ K along the $c$-axis. The susceptibility data 
from Figure\ \ref{Fig2} will be used later on to explain the ESR results.

\vspace{0.3cm}
\begin{table}
\caption{Spin-Peierls and N\'eel temperatures (when they exist) for the 
pure and Ni-doped CuGeO$_3$ samples.\label{table1}}
\begin{tabular}{lll}
$x$ & $T_{SP}\mbox{ (K)}$ & $T_N\mbox{ (K)}$
\\ \hline $0~$ & $14.25$ & $-$
\\ $0.002~$ & $13.8$ & $< 1.8$
\\ $0.0085~$ & $12.55$ & $0.6$
\\ $0.02~$ & $10.3$ & $2$
\\ $0.03~$ & $\mbox{ }8$ & $3$
\\ $0.04~$ & $-$ & $3.8$
\\ $0.052~$ & $-$ & $3.25$
\\ $0.06~$ & $-$ & $2.9$
\\
\end{tabular}
\protect\label{table1}
\end{table}

\section{ESR in \nilow -doped \cugeo}
\label{sec:V}

\subsection{Experimental results}
\label{sec:V.1}

\begin{figure}
   \begin{center}
     \mbox{\epsfxsize=8.5cm \epsffile{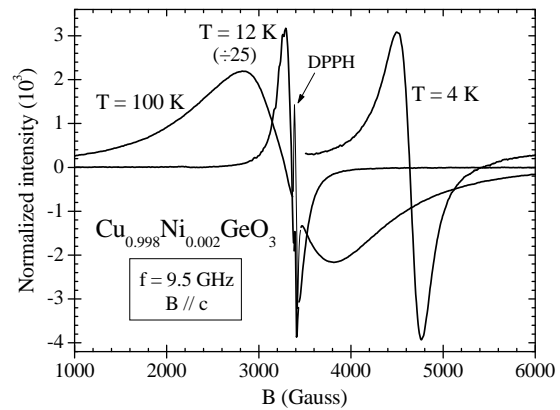}}
			\vspace{-0.3cm}
     \caption{ESR spectra obtained in X-band, with $B$$\parallel$$c$, in 
Cu$_{0.998}$Ni$_{0.002}$GeO$_3$, above and below the SP temperature. The 
figure shows the huge broadening of the ESR line on each side of $T_{SP}$ 
and the increasing shift towards high fields as $T$ decreases.}
	  \label{Fig3}
   \end{center}
\end{figure}

\begin{figure}
   \begin{center}
     \mbox{\epsfxsize=8.2 cm \epsffile{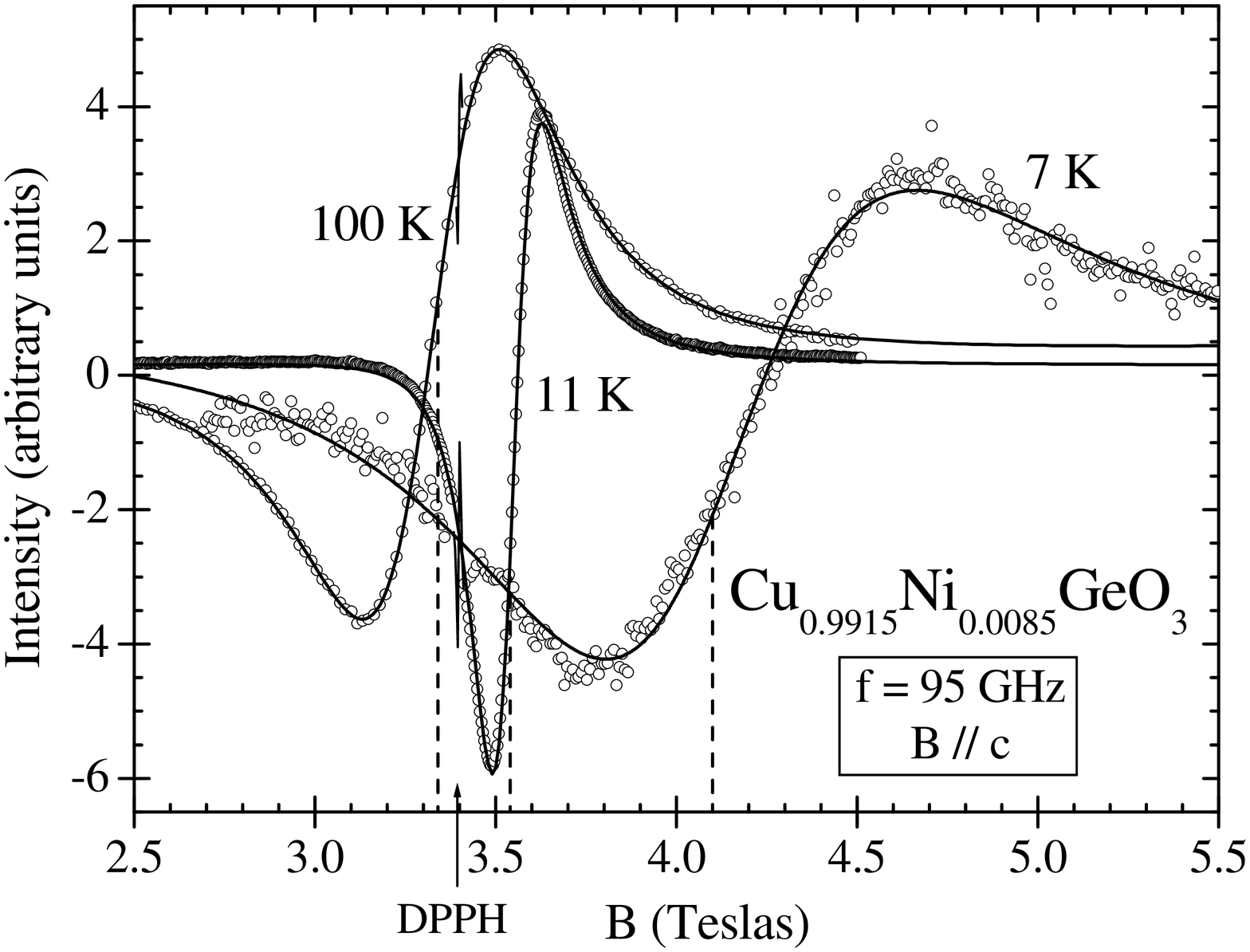}}
\vspace{-0.1cm}
     \caption{ESR spectra obtained at 95 GHz, with $B$$\parallel$$c$, in 
Cu$_{0.9915}$Ni$_{0.0085}$GeO$_3$, above and below the SP temperature. Note 
the large broadening on each side of $T_{SP}$ and the increasing shift 
towards high fields as $T$ decreases. The solid lines are the best Lorentzian fits to the experimental data (see Eq.\ (\ref{Eq1})), and the dash lines point out the position of the resonance field at each temperature, 
obtained from the fit.}
    \label{Fig4}
   \end{center}
\end{figure}

The Ni-doped CuGeO$_3$ samples were investigated at X-band along the three 
crystallographic directions and at high frequency along the $c$-axis. 
Figure\ \ref{Fig3} shows typical ESR spectra obtained in 
Cu$_{0.998}$Ni$_{0.002}$GeO$_3$ ($T_{SP}=13.8$ K) at X-band along the 
$c$-axis at $T=100$, 12 and 4 K. Figure\ \ref{Fig4} shows ESR spectra 
obtained in Cu$_{0.9915}$Ni$_{0.0085}$GeO$_3$ ($T_{SP}=12.55$ K) at 95 GHz 
along the $c$-axis at $T=100$, 11 and 7 K. In all Ni-doped compounds, a 
single lorentzian derivative ESR line was again observed but its linewidth 
$\Delta B$ and resonance field $B_{res}$ behave in a completely different 
way. Qualitatively, as compared to Fig.\ \ref{Fig1} for 1.6\% Zn-doping, 
one notices immediately on Fig.\ \ref{Fig3} for 0.2\% Ni-doping (X-band) 
and Fig.\ \ref{Fig4} for 0.85\% (95 GHz) that a $\sim 5$\% shift towards 
high field is already visible in the spectra at 12 and 11 K respectively, 
compared to those at 100 K. This shift increases to reach 24\% at 7 K (Fig.\ 
\ref{Fig4}) and 36\% at 4 K (Fig.\ \ref{Fig3}) which is much larger than 
the $\sim 1$\% shift measured in Fig.\ \ref{Fig1}. Moreover, one notes the 
large broadening induced by Ni-doping: the spectrum measured at 100 K for 
0.2\% Ni-doping (see Fig.\ \ref{Fig3}) is about three times broader than 
for 1.6\% Zn-doping (see Fig.\ \ref{Fig1}) or pure CuGeO$_3$. This large 
broadening justifies the use of higher frequencies for further studies of 
the Ni-doped CuGeO$_3$ samples.

The spectra obtained in X-band ($f=9.5$ GHz) correspond to the field 
derivative of the absorption $\chi"$ while those obtained in the 
millimetric domain ($f=95$ and 190 GHz) correspond to the field derivative 
of a combination of the absorption $\chi"$ and dispersion $\chi'$. In order 
to check the lorentzian profile and to extract precisely the integrated 
intensity (meaningfull at X band only), the linewidth, and the resonance 
field for each measured spectrum, the data obtained in the X and 
millimetric domains were fitted to the relation:

\begin{equation}
S(B)=S_0(B) + \frac{d\chi"}{dB} ~ \cos \alpha + \frac{d\chi'}{dB} ~ \sin \alpha
\label{Eq1}
\end{equation}

\vspace{-0.4cm}

\noindent where $S_0(B)$ is a linear function in $B$ accounting for the 
base line, $\alpha$ is the phase (equal to zero at X band), $\chi"(B)$ and 
$\chi'(B)$ are Lorentzian shape centered at the resonance field $B_{res}$, 
with an integrated intensity $I_{ESR}$ and a Lorentzian linewidth $\Delta 
B^{Lorentz}$. The "peak to peak" linewidth $\Delta B$ of the ESR spectra is 
given by $\Delta B = \frac{\sqrt{3}}{2} \Delta B^{Lorentz}$. Figure\ 
\ref{Fig4} illustrates the very good quality of the fits to Eq.\ 
(\ref{Eq1}) of a few ESR spectra measured in 
Cu$_{0.9915}$Ni$_{0.0085}$GeO$_3$ at 95 GHz for three different 
temperatures. Typical precisions are less than 1\% on the determination of 
$\Delta B$ and less than 0.01\% for $B_{res}$, the precision on the phase 
$\alpha$ (in the case of the millimetric range) being always better than 
1$^o$.

\subsubsection{Integrated intensity at X band}
\label{sec:V.1a}

In the case of X band, the ESR integrated intensity can be compared to the 
static susceptibility measured with the SQUID magnetometer. ESR spectra 
were measured in three \cuni samples. For $x=0.002$, the ESR line could be 
followed between 4 and 300 K, but due to the very large broadening induced 
by Ni-doping, the ESR line could be followed only up to 30 K for $x=0.0085$ 
and up to 20 K for $x=0.02$. All spectra measured for $x=0.002$ were 
analyzed using Eq.\ (\ref{Eq1}) with $\alpha=0$ and $S_0(B)=0$ and the ESR 
(twice) integrated intensity $I_{ESR}$ was obtained for each temperature. 
Due to the strong lineshift induced by Ni-doping, the resonance field must 
be taken into account and the static susceptibility measured by SQUID has 
thus to be compared with the following ratio:

\begin{equation}
I'_{ESR}=I_{ESR}~ \left ( \frac{B_0}{B_{res}} \right )
\label{Eq2}
\end{equation}

\noindent where $B_0$ is the resonance field of pure CuGeO$_3$ measured at 
the same frequency and along the same direction ({\it i.e.} corresponding 
to $g_c=2.05$ in the present case where $B$$\parallel$$c$). Note that in 
the case of non-magnetic impurities, the static susceptibility was directly 
compared to $I_{ESR}$ since the resonance field varied by less than 2\% on 
the whole temperature range.

\begin{figure}
   \begin{center}
     \mbox{\epsfxsize=8cm \epsffile{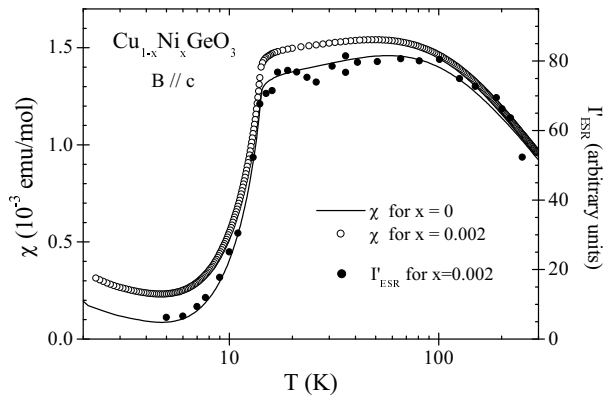}}
\vspace{0.2cm}
     \caption{Temperature dependence of the static susceptibility (SQUID) 
measured with $B=0.1$ T $\parallel$ $c$ in pure and 0.2\% Ni-doped 
CuGeO$_3$ (left scale) compared to the susceptibility derived from the ESR 
measurements ($I'_{ESR}=I_{ESR}.B_0/B_{res}$) in the same Ni-doped sample and with $B$$\parallel$$c$ (right 
scale), on a logarithmic T scale. $I'_{ESR}(T)$ fits better with the static 
susceptibility of pure CuGeO$_3$.}

    \label{Fig5}
   \end{center}
\end{figure}

Figure\ \ref{Fig5} displays the temperature dependence of $I'_{ESR}$ 
(right scale) compared to the macroscopic susceptibility vs temperature 
$\chi (T)$ in the same sample ($x=0.002$) and in pure CuGeO$_3$ (left 
scale), in a logarithmic scale. The SQUID measurements were performed in a 
0.1 T magnetic field applied along the $c$-axis while the ESR measurements 
were performed in a sweeping field from 0 to 6600 G applied along the same 
axis. By adjusting the $\chi (T)$ and $I'_{ESR}(T)$ curves at high 
temperature, one can note that the ESR integrated intensity follows better 
the susceptibility of the pure sample than that of the 0.2\% Ni-doped 
sample. This results leads one to suggest that the ESR signal would 
originate only from the Cu$^{2+}$ spins of CuGeO$_3$ while the Ni$^{2+}$ 
spins would not participate to this resonance line \cite{note1}.

\begin{figure}
   \begin{center}
     \mbox{\epsfxsize=8 cm \epsffile{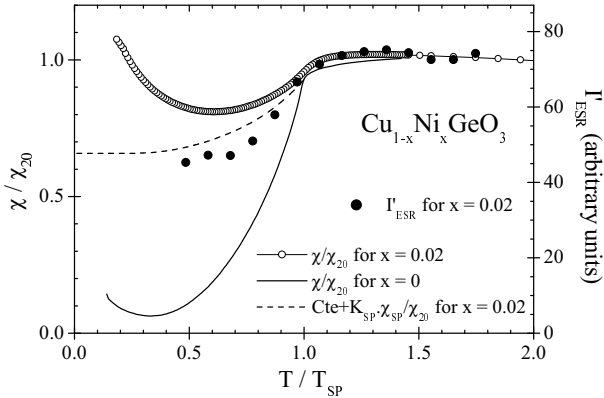}}
\vspace{0.2cm}
     \caption{Static susceptibility (SQUID) measured with $B=0.1$ T 
$\parallel$ $c$ in pure ($T_{SP}=14.25$ K) and 2\% Ni-doped CuGeO$_3$ 
($T_{SP}=10.3$ K), normalized at 20 K (left scale), compared to the 
susceptibility derived from the ESR measurements ($I'_{ESR}$) in 
the same Ni-doped sample and with $B$$\parallel$$c$ (right scale), as a 
function of the reduced temperature $T/T_{SP}$. The dash line (left scale) 
represents the SP contribution to the global susceptibility of 2\% Ni-doped 
CuGeO$_3$ arising from the model described in Ref.\protect 
\onlinecite{Grenier98b} (see text).}
    \label{Fig6}
   \end{center}
\end{figure}

To check this assumption, the same comparison was performed for a higher 
Ni-doped sample ($x=0.02$). Here again, the static susceptibility vs 
temperature curves of Cu$_{0.98}$Ni$_{0.02}$GeO$_3$ and pure CuGeO$_3$ are 
compared with the $I'_{ESR}(T)$ curve derived from the ESR measurements in 
the same Ni-doped sample. For an easier comparison, the susceptibility data 
were normalized at 20 K (by dividing for each temperature $\chi (T)$ by 
$\chi(T=20$ K)$=\chi_{20}$) and all the curves are plotted as a function of 
the reduced temperature $T/T_{SP}$, with $T_{SP}=14.25$ K for $x=0$ and 
$T_{SP}=10.3$ K for $x=0.02$ (see Fig.\ \ref{Fig6}). Indeed, for such a 
high Ni-doping level, the spin-Peierls temperatures and the absolute value 
of the static susceptibility are quite different from the pure sample (see 
Fig.\ \ref{Fig2}a). By normalizing the susceptibility and ESR data at 20 K, 
one can notice that the $I'_{ESR}$ curve lies in between the normalized 
susceptibilities of pure and Ni-doped CuGeO$_3$.

Since it was shown \cite{Grenier98b,Grenier98a} that the amplitude of the 
SP order parameter (expressed by the fraction $K_{SP}$) is reduced in 
presence of doping (by Ni, Zn, Mg, Si), we have analyzed our data 
accordingly. In particular, for \cuni with $x=0.02$, $K_{SP}$ was found to 
be equal to 0.465 \cite{Grenier98b}, {\it i.e.} only approximately one half 
of the Cu spins are able to dimerize so that the decrease of $\chi(T)$ is 
much smaller than for pure CuGeO$_3$. Moreover, the decrease of the energy 
gap upon doping further amplifies this effect. Thus, our ESR intensity data 
have to be compared with this (reduced) SP contribution instead of that of 
the full susceptibility of pure CuGeO$_3$. The dash line from Fig.\ 
\ref{Fig6} shows the temperature dependence of this SP contribution 
(divided by the Cu$_{0.98}$Ni$_{0.02}$GeO$_3$ susceptibility at 20 K) 
normalized at $T/T_{SP}=1$ with the curve of the pure sample. One can note 
the good agreement between this curve and the $I'_{ESR}$ data. This 
analysis further supports the idea that the ESR signal only comes from the 
Cu$^{2+}$ spins. We will come back to these results when discussing the 
lineshift results (see Sec.\ \ref{sec:V.2}). The same study has been 
performed for a magnetic field applied along the $a$-axis and the results 
were found to be similar.

\subsubsection{Linewidth}
\label{sec:V.1b}

\begin{figure}
   \begin{center}
     \mbox{\epsfxsize=8cm \epsffile{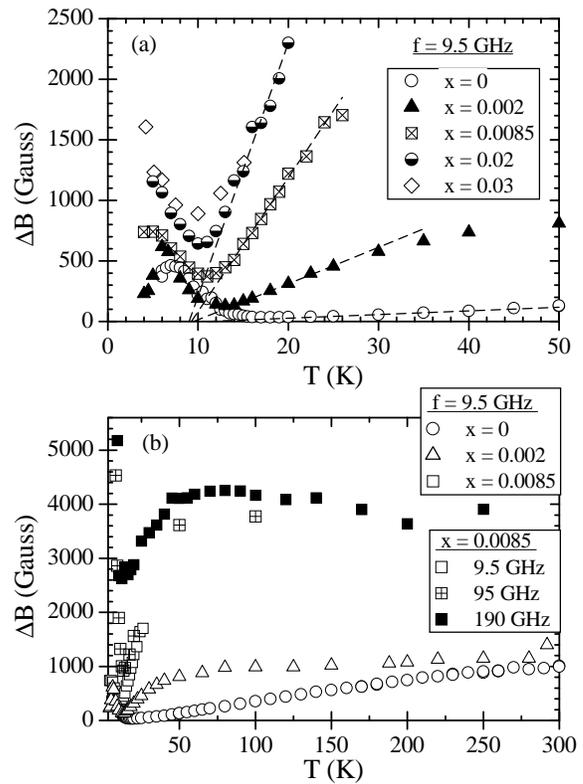}}
\vspace{0.2cm}
     \caption{Temperature dependence of the ESR linewidth measured with 
$B$$\parallel$$c$ in pure and Ni-doped CuGeO$_3$ for various 
concentrations: a) Low temperature part at X band; b) High temperature part 
at X band and higher frequencies. Note the plateau above 50 K for 0.2 and 
0.85\% Ni.}
    \label{Fig7}
   \end{center}
\end{figure}

Besides the large ESR shift, the other main qualitative effect of Ni-doping is a very large broadening of the 
ESR line as compared to pure CuGeO$_3$. This is further contrasted by the 
linewidth data of CuGeO$_3$ doped by the non-magnetic impurities Zn, Mg or 
Si \cite{Grenier99}. In Fig.\ \ref{Fig7}a, the low temperature linewidth 
data are presented at X band for pure CuGeO$_3$ and various Ni-doping 
levels. It appears that above $T_{SP}$ and up to about 40 K, the linewidth 
increases linearly with temperature and is approximately proportional to 
the Ni concentration at a rate of $\Delta B / x \sim 1000$ G/\%Ni for 
$T=20$ K and a slope $\Delta B / (x.T) \sim 100$ G.K$^{-1}$/\%Ni.

However, as shown in Fig.\ \ref{Fig7}b, contrary to the pure CuGeO$_3$ 
linewidth data, where the linear temperature dependence extends almost up 
to room temperature (see also Ref. \onlinecite{Oseroff94,Yamada96}), the Ni-doped 
CuGeO$_3$ ESR linewidth reaches a plateau above 50 K independent of both 
the temperature and the magnetic field. This plateau is proportional to the 
Ni concentration at a rate $\Delta B_{plateau} / x \simeq 4600$ G/\%Ni. At 
temperatures lower than $T_{SP}$, a temperature behavior much closer to 
that of pure CuGeO$_3$ is observed with a comparable rate of broadening (see 
Fig.\ \ref{Fig7}a), but now strongly increasing with field (see Fig.\ 
\ref{Fig7}b). Whereas we cannot account at present for the temperature dependence of the linewidth, the order of magnitude of $\Delta B_{plateau}$ is discussed in Sec.\ \ref{sec:V.2}. 

\subsubsection{Lineshift along the c-axis}
\label{sec:V.1c}

We now focus on the ESR lineshift $\delta B=B_{res}-B_0$, measured for 
$B$$\parallel$$c$, $B_{res}$ being the resonance field of the studied sample 
and $B_0$ the resonance field of pure CuGeO$_3$ (corresponding to 
$g_c=2.05$) measured at the same frequency. The aim of the following 
analysis is to study the field (or frequency), concentration and temperature dependence of 
the shift in order to derive experimentally an expression of $\delta 
B(f,x,T)$.

\begin{figure}
   \begin{center}
     \mbox{\epsfxsize=7.5cm \epsffile{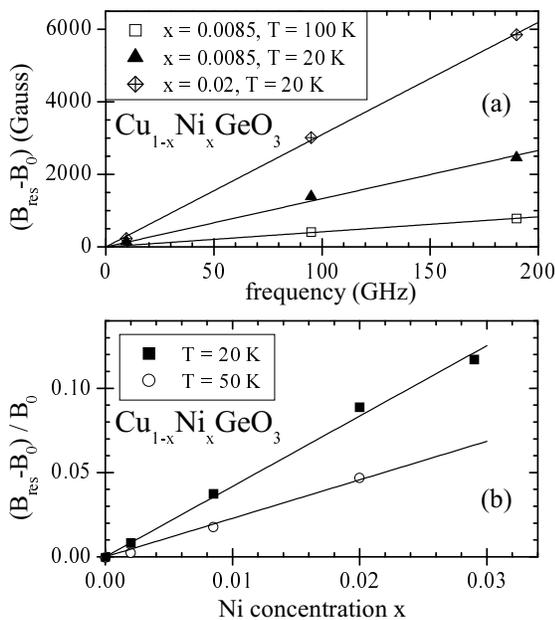}}
\vspace{0.2cm}
     \caption{(a) Frequency dependence of the ESR lineshift $\delta B = B_{res} - B_0$ measured in 
Cu$_{1-x}$Ni$_{x}$GeO$_3$ at two given temperatures. (b) Evolution with the 
nickel concentration of the relative lineshift measured at two different 
temperatures in four Cu$_{1-x}$Ni$_{x}$GeO$_3$ single crystals. These two 
figures establish the proportionality (a) between the lineshift and the 
frequency at given temperature and Ni concentration, and (b) between the 
relative shift and the Ni concentration, at a given temperature (see solid 
lines).}
    \label{Fig8}
   \end{center}
\end{figure}

Figure\ \ref{Fig8}a shows the frequency dependence of the lineshift 
measured for two different samples, 0.85 and 2\% Ni-doped CuGeO$_3$, and 
two different temperatures. This figure shows that the shift is 
proportional to the working frequency (see solid lines) and thus to $B_0$, 
which proves that it is a genuine "$g$-shift". This shift increases with $x$ 
and $T^{-1}$. To study the concentration and temperature dependences, we 
will thus now refer to the frequency-independant relative shift $\delta B/B_0$. Figure\ \ref{Fig8}b shows the Ni 
concentration dependence of $\delta B/B_0$ at two given temperatures, 
$T=20$ and 50 K (data from various frequencies were combined together to 
obtain this figure). For both temperatures, the relative shift is 
proportional to the Ni concentration, up to 3\% (see solid lines). At this 
stage, we thus have the relation:

\begin{equation}
\frac{\delta B}{B_0}(x,T)=C(T).x
\label{Eq3}
\end{equation}

\begin{figure}
   \begin{center}
     \mbox{\epsfxsize=8.5cm \epsffile{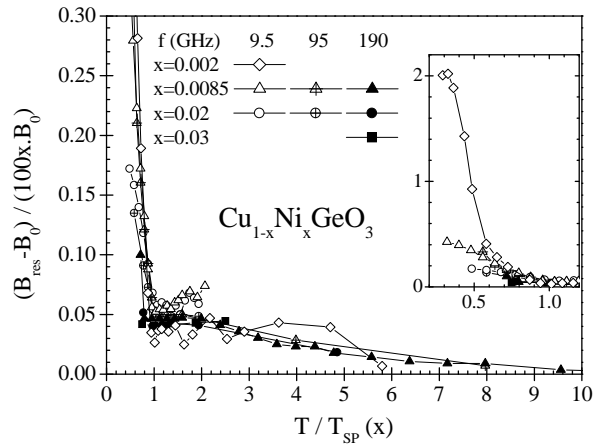}}
\vspace{0.2cm}
     \caption{Relative lineshift $\delta B / B_0$ normalized by the Ni 
concentration (divided by $100 x$) as a function of the reduced temperature 
$T/T_{SP}(x)$ for all studied Ni-doped CuGeO$_3$ samples (various 
concentrations at various frequencies). The inset shows the low temperature part.}
    \label{Fig9}
   \end{center}
\end{figure}

All our experimental results about the lineshift are gathered in Figure\ 
\ref{Fig9} using the above relation, {\it i.e.} the relative lineshift 
normalized to the Ni concentration $\delta B / (100 x . B_0)$ is plotted as 
a function of the reduced temperature $T/T_{SP}(x)$. This figure shows the 
good agreement of relation (\ref{Eq3}) with our data. All the experimental 
points are roughly on a unique curve $C(T)$, at least above $T_{SP}$. Below 
the spin-Peierls temperature, the temperature dependence of $\delta B/B_0$ 
depends on the Ni-doping level. Indeed, the data at 9.5 and 95 GHz are well 
superimposed down to 0.7 $T_{SP}$ for both samples with $x=0.0085$ and 
$x=0.02$, but the increase at lower temperature is stronger as the Ni 
concentration decreases (see inset of Fig.\ \ref{Fig9}).

\begin{figure}
   \begin{center}
     \mbox{\epsfxsize=7.5cm \epsffile{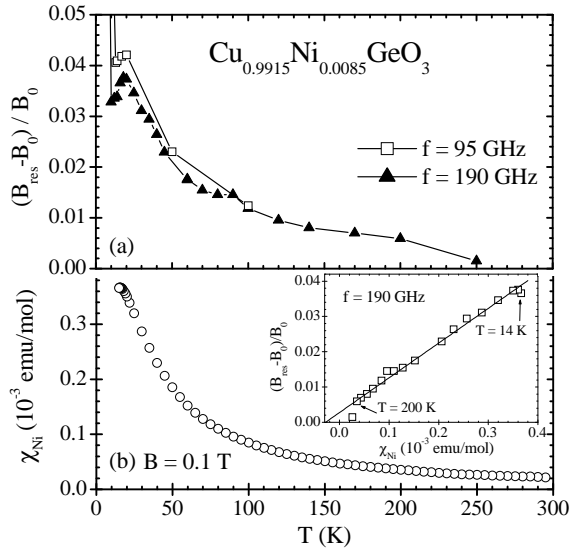}}
		\vspace{0.2cm}
     \caption{Temperature dependence in the paramagnetic phase of (a) the 
relative ESR lineshift and (b) the static susceptibility of the Ni ions, 
measured in Cu$_{0.9915}$Ni$_{0.0085}$GeO$_3$. The inset of Fig.\ \ref{Fig10}b 
illustrates the relative ESR lineshift as a function of the Ni 
susceptibility, at each temperature.}
    \label{Fig10}
   \end{center}
\end{figure}

\begin{figure}
   \begin{center}
     \mbox{\epsfxsize=7.5cm \epsffile{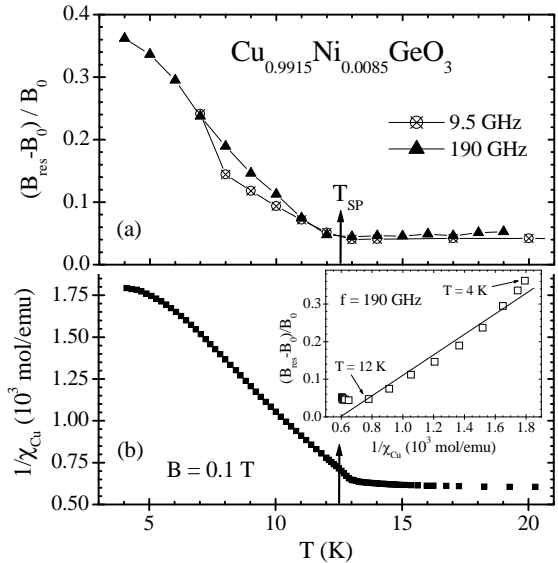}}
		\vspace{0.2cm}
     \caption{Temperature dependence in the spin-Peierls phase of (a) the 
relative ESR lineshift and (b) the inverse static susceptibility of the Cu 
dimers, measured in Cu$_{0.9915}$Ni$_{0.0085}$GeO$_3$. The inset of Fig.\ 
\ref{Fig11}b illustrates the relative ESR lineshift as a function of the 
inverse Cu susceptibility, at each temperature.}
    \label{Fig11}
   \end{center}
\end{figure}

Thus, in order to study in detail the temperature dependence of the shift, 
we consider separately the paramagnetic (P) and SP phases. The temperature 
dependence of the relative shift in the 0.85\% Ni-doped sample is 
illustrated for two different frequencies in Figures\ \ref{Fig10}a for 
$T>T_{SP}$ and\ \ref{Fig11}a for $T<T_{SP}$. Between 300 and 20 K, $\delta 
B/B_0$ increases up to 4\% (see Fig.\ \ref{Fig10}a) and its temperature 
dependence behaves like the nickel susceptibility $\chi_{Ni}(T)$ (see Fig.\ 
\ref{Fig10}b). The latter has been derived by subtracting the susceptibility 
of pure CuGeO$_3$ to that of the Ni-doped sample (both presented in Fig.\ 
\ref{Fig1}b). The inset of Figure\ \ref{Fig10}b shows that $\delta B(T)/B_0$ 
is strictly proportional to $\chi_{Ni}(T)$ in the P phase.

\vspace{0.2cm}

\vspace{-0.3cm}

Concerning the low temperature part, one can notice a strong enhancement of 
the line shift below 12.5 K at 9.5 and 190 GHz (see Fig.\ \ref{Fig11}a). 
This temperature corresponds to the SP transition, thus this behavior is 
mainly related to the dimerization of the Cu ions into $S=0$ states. 
Figure\ \ref{Fig11}b illustrates the temperature dependence of the inverse 
of the Cu-dimers susceptibility (spin-Peierls contribution in the triplet 
state $S=1$) $\chi_{Cu}(T)$, which was obtained by subtracting a 
Curie-Weiss contribution (coming from the Ni ions) to the total 
susceptibility\cite{Grenier98b}. As can be seen in the inset of Figure\ 
\ref{Fig11}b, $\delta B(T)/B_0(T)$ varies linearly with $1/\chi_{Cu}(T)$. 
This analysis of the data leads to the following relation for the relative 
shift:

\begin{equation}
\frac{\delta B}{B_0}(x,T)\propto x.\frac{\chi_{Ni}(T)}{\chi_{Cu}(T)}
\label{Eq4}
\end{equation}

For the same reason as for the integrated intensities (decrease of the fraction $K_{SP}$ when x increases), this formula is indeed consistent with the inset of Fig.\ \ref{Fig9} where the relative shift below $T_{SP}$ shows a fourfold increase as the Ni doping level decreases from 0.85\% to 0.2\%. 

\subsubsection{Anisotropy of the lineshift}
\label{sec:V.1d}

\begin{figure}
   \begin{center}
     \mbox{\epsfxsize=7.5cm \epsffile{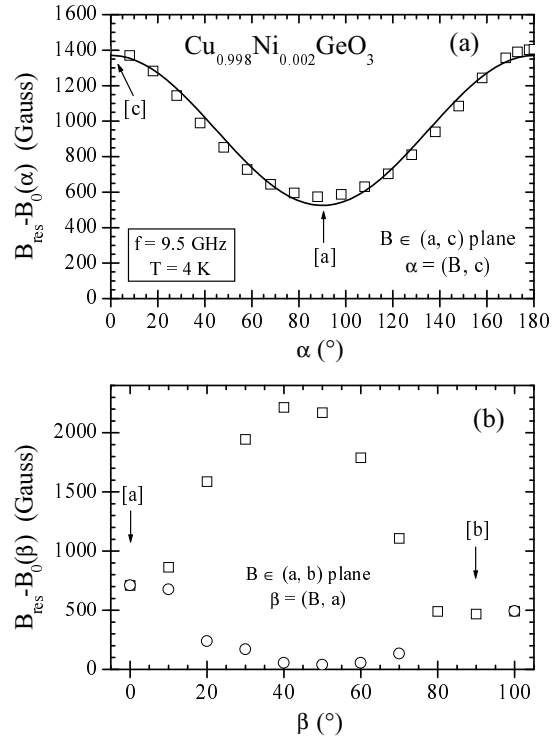}}
		\vspace{0.2cm}
     \caption{Anisotropy of the ESR lineshift in the $(a,c)$ and $(a,b)$ 
planes measured in X-band, at $T=4$ K in 0.2\% Ni-doped CuGeO$_3$. The open 
squares correspond to the afore mentioned ESR line and the open circles to 
the new line that appears in the $(a,b)$ plane.}
    \label{Fig12}
   \end{center}
\end{figure}

The anisotropy of the lineshift has also been carefully studied at X-band 
in the 0.2\% Ni-doped compound at 4 K, in the $(a,c)$ and $(a,b)$ planes 
(see Fig.\ \ref{Fig12}).

The shift $\delta B$ was found to be very 
anisotropic in both planes and a second, less intense, resonance line was 
observed in the $(a,b)$ plane. In the $(a,c)$ plane, the shift is maximum along $c$ ($\simeq$ 1400 Gauss) and 
minimum along $b$ ($\simeq$ 600 Gauss), following $\delta B (\alpha) = \delta B_{ac} 
+ \delta B_{anis} \cos \alpha$ with $\delta B_{ac} \simeq 950$ and $\delta 
B_{anis} \simeq 420$ Gauss. In the $(a,b)$ plane, the shift is minimum and 
equivalent along the $a$ and $b$ axes while a maximum shift (2200 Gauss) is 
found at $\beta \simeq 45^o$ (where $\beta$ is the angle between the 
applied magnetic field and the $a$-axis) possibly reflecting the anisotropy 
of the Cu to the Ni site coupling. Note that the anisotropy of the 
lineshift along the crystal axes follows that of the static susceptibility 
(see Fig.\ \ref{Fig2}b for the 6\% Ni-doped CuGeO$_3$ sample).

\subsection{Resonance Model}
\label{sec:V.2}

The model used to interpret the ESR data of CuGeO$_3$ doped with Ni at very 
low concentrations is constrained by the
above mentioned experimental observations which we briefly recall:

i- At all temperatures, along the $a$, $b$, and $c$ directions, only one Lorentzian shaped (except at the lowest temperature) ESR line is 
observed, above and below the spin-Peierls transition, which we attribute 
to the paramagnetic Cu$^{2+}$ spins above $T_{SP}$ and to the excited 
triplet Cu$^{2+}$ spins below $T_{SP}$.

ii- No resonance consistent with Ni$^{2+}$ in a distorted octahedral site 
is observed in the temperature range $4<T<300$ K, neither at 9.5 GHz nor up 
to 190 GHz and 10 Tesla.

iii- The ESR integrated intensity, normalized to the resonance field, does 
not scale with the static susceptibility measured in the same Ni-doped 
sample, but only to the proportion of Cu spins being in the SP triplet 
state, which supports i-.

iv- The width of the Cu$^{2+}$ resonance strongly increases with the Ni 
concentration (at a rate of 0.1 Tesla/\% Ni at 20 K).

v- A resonance shift of the Cu$^{2+}$ line towards higher fields, 
proportional to the concentration of the Ni$^{2+}$ ions, is observed, above 
and below the spin-Peierls transition. This shift is proportional to the Ni 
ions susceptibility and inversely proportional to the Cu ions 
susceptibility. This shift thus becomes particularly noticeable below the 
SP transition. The shift is anisotropic, being roughly twice as large for 
the $c$-direction as for the $a$ and $b$ directions, which are about 
equivalent: it appears then that this shift is not related to the 
$g-$factor anisotropy of the pure CuGeO$_3$ system ($g_a=2.15$, $g_b=2.26$, 
$g_c=2.05$).

A very detailed investigation of a two coupled paramagnetic spins system,
present in RbMnF$_3$ doped with Ni, Co or Fe, has been performed by Gulley
and Jaccarino\cite{Gulley}. For 1 to 5\% Ni-doping, contrary to the present
work, no discernable linewidth broadening together with a small $g$-shift of
0.5\% are observed. This situation was quantitatively analyzed in terms of
very strongly exchange-coupled Ni and Mn spins. This is clearly not the case
for CuGeO$_3$ : Ni but does apply to the case of CuGeO$_3$ : Zn, Mg or Si. Indeed, in addition to the absence of broadening and to the 1\% $g$-shift, the fact that the ESR integrated intensity fits well the static susceptibility constitutes a compelling evidence for the strong coupling between the bulk Cu$^{2+}$ spins and the Cu$^{2+}$ spins induced by the non magnetic impurities.

The model used for our interpretation is an extension of Kittel's analysis 
for the ferrimagnetic resonance shift in the presence of strongly relaxing 
rare earth dopants \cite{Kittel59}. Indeed the coupling $J_c\simeq 120-180$ 
K of the Cu$^{2+}$ spins along each of the two (non equivalent 
\cite{Pilawa97}) 1D chains of CuGeO$_3$ is strong enough to ensure a 
single, exchange-narrowed mode in presence of an antiferromagnetic coupling 
$J_N$ (induced on its two Cu$^{2+}$ first nearest neighbors) by a Ni$^{2+}$ 
$S=1$ spin. In order to account for our observation of a single resonance 
line, we are forced to assume that (i) either the relaxation rate of the 
Ni$^{2+}$ spin is sufficiently large, compared to $J_N$, to decouple the 
Ni$^{2+}$ spins from the precession of the Cu$^{2+}$ spins or that (ii) 
this decoupling occurs through a sufficiently large zero-field splitting 
removing the Ni$^{2+}$ resonance away from the Cu$^{2+}$ resonance. In that 
case, we can modelize the spin dynamics of CuGeO$_3$:Ni by two coupled 
Bloch-like equations, one for the tightly coupled together Cu$^{2+}$ spin 
system (magnetization $\vec{M}$) and one for the fast relaxing Ni$^{2+}$ 
spins (magnetization $\vec{N}$). For sake of simplicity we will assume that 
the gyromagnetic ratio $\gamma_{Cu}\simeq \gamma_{Ni} = \gamma$ so that the 
decoupling occurs only through the relaxation rate of the Ni$^{2+}$ spins. 
The coupled equations are then written with the notation: $T_{Ml}^{-1}$ the 
Cu spin-lattice relaxation rate, $T_{Nl}^{-1}$ the Ni spin-lattice 
relaxation rate, $T_{MN}$ and $T_{NM}$ the Cu/Ni spin-spin cross relaxation 
rates between Cu and Ni (related by $T_{MN}/T_{NM}=N^0/M^0$, {\it i.e.} the 
ratio of the equilibrium magnetizations, by detailed balance 
\cite{Barnes81}).

\begin{eqnarray}
\frac{d\vec{M}}{dt}&=&\gamma \vec{M} \times (\vec{H}+\lambda 
\vec{N})-\frac{\vec{M}}{T_{MN}}+\frac{\vec{N}}{T_{NM}}-\frac{\vec{M}-\vec{M} 
^0}{T_{Ml}}\\
\frac{d\vec{N}}{dt}&=&\gamma \vec{N} \times (\vec{H}+\lambda 
\vec{M})-\frac{\vec{N}}{T_{NM}}+\frac{\vec{M}}{T_{MN}}-\frac{\vec{N}-\vec{N} 
^0}{T_{Nl}}
\label{Eq5-6}
\end{eqnarray}

Solving for the transverse components as usual leads to the determinant:

\begin{equation}
\left|
\begin{array}{cc}
\Omega_N & \lambda_N \\ \lambda_M & \Omega_M
\end{array}
\right| =0
  \label{Eq7}
\end{equation}

with
\begin{eqnarray}
\Omega_M &=& \omega - \omega_0 - iT_{Ml}^{-1}-\lambda_M\\
\Omega_N &=& \omega - \omega_0 - iT_{Nl}^{-1}-\lambda_N\\
\lambda_M &=& \gamma \lambda M^0 + i T_{NM}^{-1}\\
\lambda_N &=& \gamma \lambda N^0 + i T_{MN}^{-1}
\label{Eq8-11}
\end{eqnarray}

\noindent where $\omega_0$ is the unshifted resonance frequency. In the 
limit of $T_{Nl}^{-1}\rightarrow \infty$ we get \cite{note2}:

\begin{equation}
\omega - \omega_0 = \lambda \gamma N^0 + i(T_{MN}^{-1}+T_{Ml}^{-1})
\label{Eq12}
\end{equation}

This yields indeed a Cu$^{2+}$ resonance shifted to higher fields if 
$\lambda <0$ {\it i.e.} an antiferromagnetic coupling from Ni$^{2+}$ to 
Cu$^{2+}$, and proportional to the Ni$^{2+}$ static magnetization $N^0$. 
However, a major difference appears between our CuGeO$_3$:Ni model and 
Kittel's two sublattice model. In the classical sublattice spin dynamics 
the coupling constant $\lambda$ represents the exchange energy density, 
through $-\lambda \vec{M}_A.\vec{M}_B$, meaning that every spin A is 
coupled to every spin B. In our case, we want to specify that the Ni spin 
is coupled to its two first nearest neighbors Cu$^{2+}$ by $J_N$ and that 
the effective coupling $\lambda$ is the resulting average of the Cu-Ni 
exchange $J_N$ among the Cu$^{2+}$ spins (tightly coupled through $J_c$). 
This average coupling $\lambda$ occurs however only from those Cu$^{2+}$ 
spins which are not condensed into singlets $S=0$ below the spin Peierls 
transition. Writing that the exchange energy density is identical in both 
descriptions yields:

\begin{equation}
\lambda = \frac{2J_N}{\hbar^2 \gamma^2} \frac{1}{(1-n_{SP})\mathcal{N}}
\label{Eq13}
\end{equation}
where $n_{SP}$ is the fraction of condensed singlets spins {\it i.e.} the 
spin-Peierls order parameter and $\mathcal{N}$ is the number of Cu ions per 
unit volume. The shift is then given by:

\begin{equation}
\omega - \omega_0 = \frac{2J_N}{\hbar^2 \gamma} \frac{N^0}{(1-n_{SP}) 
\mathcal{N}}
\label{Eq14}
\end{equation}

Thus, Eq.(\ref{Eq14}) indeed predicts the observed dependence for the 
lineshift (see Eq.(\ref{Eq4})). The merit of this formulation is that we 
use the same argument above and below $T_{SP}$. The above equation can be 
rewritten (in the P phase) in terms of our measured quantities to yield 
$J_N$ by:

\begin{equation}
J_N = \frac{1}{2} (g\mu_B)^2 N_A \left [ \frac{\delta B/B_0}{\chi^0_{Ni}} 
\right ]
\label{Eq15}
\end{equation}

The term between brackets is depicted in the inset of Fig.\ \ref{Fig10}b. We 
find $J_N = 75 \pm 10$ K where the error is estimated from the spread of 
the data at various $x$ and frequencies. This value for $J_N$ was checked 
to be quite consistent with the weak coupling approximation. However this 
value is three times larger than the one obtained above (\S IV) from the 
fit to the static susceptibility. We think that the value of $J_N$ derived 
from the ESR data is closer to the truth since the zero field splitting of 
the Ni$^{2+}$ ion was not incorporated in the susceptibility analysis: This 
parameter is a priori not required in the ESR model since Ni$^{2+}$ is 
considered as "non resonant".

As concerns the linewidth, a direct application of the weak coupling exchange
broadening calculated by Gulley and Jaccarino (equation (2.12) from Ref. \onlinecite{Gulley})
with $J = J_c \simeq$ 150 K and $J'=J_N\simeq$ 75 K yields a temperature
independant broadening of 8000 G/\% Ni, to be compared with 4600 G/\% Ni
observed between 50 and 300K.

\section{Antiferromagnetic resonance in \nilow -doped \cugeo}
\label{sec:VI}

\begin{figure}
		\begin{center}
 		\mbox{\epsfxsize=7.2 cm \epsffile{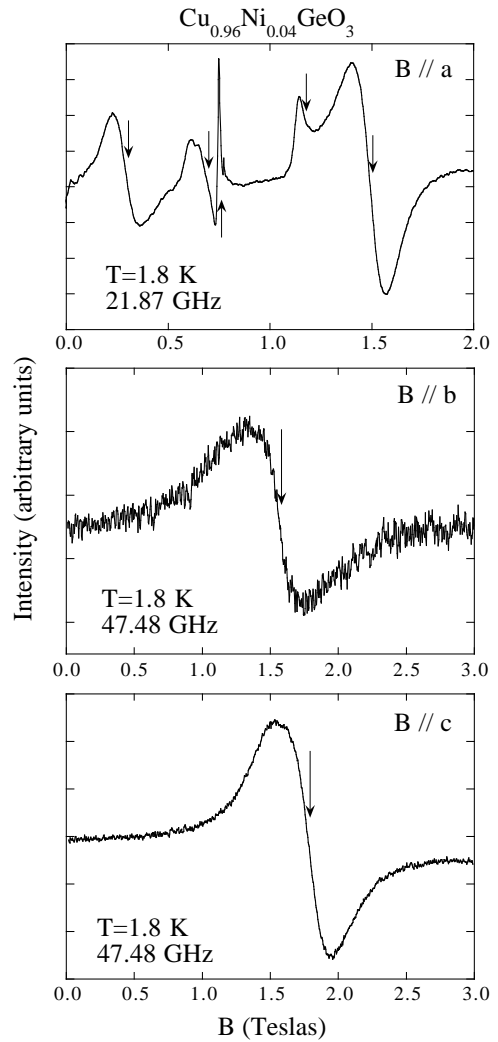}}
		\vspace{0.2cm}
    	\caption{Typical ESR signals in a single crystal of 
Cu$_{0.96}$Ni$_{0.04}$GeO$_3$ at 1.8 K for the three crystallographic 
orientations.}
		\label{Fig13}
		\end{center}
\end{figure}

Typical AFMR signals measured in the $x=0.04$ sample at 1.8 K in a field applied along the 
$a$, $b$, and $c$ axes are shown in Fig.\ \ref{Fig13}. Five ESR lines were 
observed at 21.87 GHz for $B$$\parallel$$a$. The lines at 0.3, 1.2 and 1.5 T 
correspond to AFMR lines. The line at 0.75 T originates from a standard 
DPPH (1,1-diphenyl-2-picrylhydrazyl) sample. The line at 0.7 T probably 
corresponds to an Electron Paramagnetic Resonance (EPR) signal from the 
sample. For $B$$\parallel$$b$ and $B$$\parallel$$c$, one ESR line, which 
corresponds to an AFMR signal, was observed around 47.5 GHz. The EPR signal 
which was observed along the $a$-axis in the AF ordered state was not 
observed for these two field directions\cite{note3}. This is consistent with the 
recent theory \cite{OM00} which predicts the existence of an EPR signal 
even in the ordered phase when the external field is applied parallel to 
the easy axis.

From the AFMR results, we can identify the easy, second-easy, and hard axes 
as the $a$, $c$, and $b$ axes, respectively, as will be shown below. The 
angular dependence of the resonance fields at X band was also 
measured. Resonance points around 0.9 T show a minimum when the field is
parallel to the $a$-axis. From this result we confirmed that the spins 
point along the $a$-axis within an experimental error of $\sim 2^o$, which 
had not clearly been determined by the susceptibility \cite{Koide96} and 
the neutron scattering measurements \cite{Coad96}. We also performed AFMR 
measurements in the $x=0.03$ sample at $T=1.8$ K. The AFMR signals are less 
clear mainly due to the lower $T_N$. However, it was found that the 
magnetic anisotropy is the same as in the $x=0.04$ sample.

\begin{figure}
	\begin{center}
		\mbox{\epsfxsize=8.5cm \epsffile{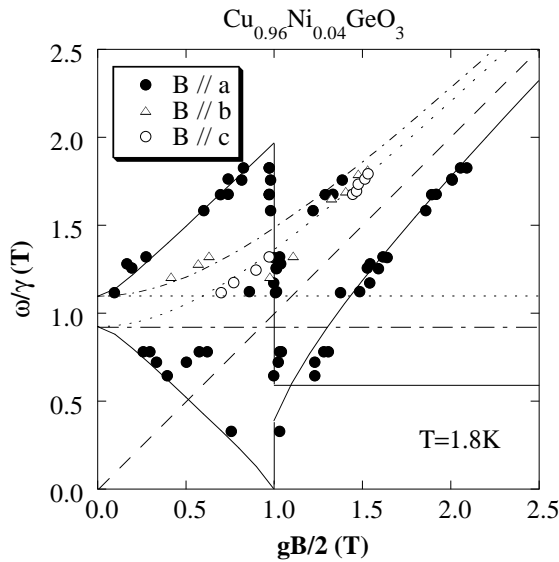}}
		\vspace{0.2cm}
		\caption{The frequency vs magnetic-field diagram of the resonance points 
at 1.8 K in a Cu$_{0.96}$Ni$_{0.04}$GeO$_3$ single crystal. Closed circles, 
open triangles, and open circles denote experimental data and solid, 
dash-dotted, and dotted curves are theoretical ones for $B$$\parallel$$a,b$, 
and $c$, respectively.}
   \label{Fig14}
  \end{center}
\end{figure}

Figure\ \ref{Fig14} shows the frequency vs magnetic-field plot of the 
resonance points at 1.8 K. Here, the angular frequency ($\omega$) is 
divided by $\gamma \equiv 2\pi g\mu_{\rm B}/h$ ($\mu_{\rm B}$: Bohr 
magneton, $h$: Planck's constant) to express it in magnetic field units and 
$B$ is scaled by the $g$ value for the respective field directions. The 
$g$ values we used are, $g_a=1.72$, $g_b=1.80$, and $g_c=1.65$, which are 
80$\%$ of the $g$ values in pure CuGeO$_3$ and are also consistent with 
those obtained from the ESR measurements at $\sim5$ K. 

\vspace{-0.6cm}
\begin{table}
\caption{Observed $C_1$, $C_2$, and $\alpha$ in CuGeO$_3$
doped with Si, Zn, and Ni.\label{table2}}
\begin{tabular}{lcccc}
$\rm impurity$&$T_N$ (K)&$C_1$ (T$^2$)&$C_2$ (T$^2$)&
$\alpha$\\
\tableline
Si (2$\%$)\tablenote{Ref. \protect\onlinecite {Nojiri97}}&4.8&1.3$\pm$0.1&3.8$\pm$0.2
&0.78$\pm$0.06\\
Zn (4$\%$)\tablenote{Ref. \protect\onlinecite {Hase96}}&4.2&0.89&2.1&0.85\\
Ni (4$\%$)&4.0&0.85&1.2&0.90\\
\end{tabular}
\protect\label{table2}
\end{table}

The data were analyzed with the AFMR theory for a two-sublattice antiferromagnet with 
orthorhombic anisotropy \cite{Nagamiya55}. This theory was used in analyzing 
Zn- \cite{Hase96,Fronzes97} and Si-doped samples \cite{Nojiri97}, and is conveniently reviewed in Ref. \onlinecite{KKatsumata}. We 
obtained the parameters $C_1$(=$2AK_1$) = 0.85 (T$^2$), $C_2$(=$2AK_2$) = 1.2 
(T$^2$), and $\alpha$(=$1-\chi_\parallel/\chi_\perp$) = 0.90, where $A$, 
$K_i$ ($i=1$ and 2), $\chi_\parallel$, and $\chi_\perp$ are the molecular 
field coefficient, the anisotropy constants, the susceptibilities along the 
easy axis and perpendicular to it, respectively. The lines in Fig.\ 
\ref{Fig14} are fits with the classical AFMR theory. The parameter $\alpha$ 
obtained with AFMR measurements is similar to those in 
Cu$_{0.96}$Zn$_{0.04}$GeO$_3$ \cite{Hase96,Fronzes97} and 
CuGe$_{0.98}$Si$_{0.02}$O$_3$ \cite{Nojiri97} as shown in Table\ \ref{table2} but is 
much larger than that obtained from the magnetic susceptibility 
measurements ($\alpha \sim 0.35$). The susceptibility parallel to the $a$-axis 
of Cu$_{0.96}$Ni$_{0.04}$GeO$_3$ does not go to zero with decreasing 
temperature below $T_N$ as shown in Fig.\ \ref{Fig2}a. This is quite different from 
what a classical theory of antiferromagnetism predicts for 
$\chi_\parallel$. We found that the measured $\chi_a$ can be interpreted 
as a sum of a paramagnetic susceptibility which obeys the Curie law and an 
antiferromagnetic one. The amount of the paramagnetic "centers" is 
estimated to be about 1\% of the total spins assuming $S=\frac{1}{2}$ and 
$g=2.1$. The antiferromagnetic susceptibility decreases with decreasing temperature more 
rapidly than $\chi_a$, thus giving a larger value for $\alpha$.

We now discuss the origin of the magnetic anisotropy in Ni-doped CuGeO$_3$. 
In order to explain this, we consider three terms; single ion anisotropy 
($E_{SI}$), dipole-dipole ($E_D$), and anisotropic exchange ($E_{AE}$) 
energies. Moriya and Yosida discussed theoretically the origin of the 
anisotropy in CuCl$_2\cdot$2H$_2$O.\cite{MY53} They considered 
both of $E_D$ and $E_{AE}$ terms (note that no $E_{SI}$ term is possible for 
Cu$^{2+}$ spin). They estimated $\mid$$E_D$$\mid \sim 10^{-4}$ cm$^{-1}$/ion and 
$\mid$$E_{AE}$$\mid \sim 10^{-3}$ cm$^{-1}$/ion, respectively. 
Although the crystal structure of CuGeO$_3$ is different from 
that of CuCl$_2\cdot$2H$_2$O, we expect that the value of $E_D$ in 
CuGeO$_3$ is not very much different from the one given above. For 
$E_{AE}$ in CuGeO$_3$, we estimate a ten times larger value, because the 
exchange interaction is about ten times larger than for CuCl$_2\cdot$2H$_2$O. 
On the other hand, $\mid$$E_{SI}$$\mid \sim 1$ cm$^{-1}$/ion for 
Ni$^{2+}$ (Ref. \onlinecite{Abragam}) so that even a small amount of doping will change 
drastically the anisotropy of the system. The $g$ value and the single ion 
anisotropy constant, $D$, for Ni$^{2+}$ in an axially distorted octahedral 
environment are related to the spin-orbit coupling and to the crystal field 
levels. In the well documented case of trigonal elongation of the octahedron, as in CsNiF$_3$\cite{Dupas}, this distorsion leads to $D>0$ (and g$_{\perp}$(Ni)$ > g_{\parallel}$(Ni)). In contrast, in the case of CuGeO$_3$, one expects an elongation to yield $D < 0$ similar to the Tutton salts\cite{Griffiths} and hence give rise to an easy local axis. Here, we denote the single ion anisotropy term as $DS_z^2$.

The easy axis for Ni$^{2+}$ spin in CuGeO$_3$ lies in the $ab$ plane and is 
directed alternately when one moves from one site to the other along the 
$b$-axis. Because of the antiferromagnetic interaction, the spins will 
point either close to the $a$-axis or the $b$-axis with antiferromagnetic 
arrangement along the $b$-axis, the former being favoured by $E_D$. Thus, 
the mean easy axis is the $a$-axis. Then the easy and second easy axes in 
Zn- and Si-doped samples are interchanged in the Ni-doped sample.

The $C_1$ and $C_2$ values for various impurity-doped CuGeO$_3$ are 
summarized in Table\ \ref{table2}. $C_1$ and $C_2$ correspond to the geometric mean of 
the exchange field and anisotropy fields along the second-easy and the hard 
axes, respectively. It is noted that the two parameters are very close in 
the Ni-doped sample. The parameter $C_2$ is related to the anisotropy when 
one rotates the magnetic moments from the $c$ to $b$ axes in CuGeO$_3$:Zn 
and from the $a$ to $b$ axes in CuGeO$_3$:Ni. In the Zn-doped 
sample, one loses both of $E_D$ and $E_{AE}$ for this rotation, 
while in the Ni-doped sample, $E_D$ and a part of $E_{AE}$ are lost 
($E_{AE}$ in the $ab$ plane is less anisotropic because $g_a$ and $g_b$ 
are nearly equal). So, $C_2$ in the Ni-doped sample is smaller than that 
in the Zn-doped sample. Consequently, $C_1$ and $C_2$ become closer in the
CuGeO$_3$:Ni sample.

\section{Conclusion}
\label{sec:VII}

We have presented in this paper a systematic study of the spin resonance 
properties of CuGeO$_3$ in the paramagnetic, spin-Peierls and 
antiferromagnetic regimes, when doped with Ni, and compared these new 
results with previously obtained ESR data on CuGeO$_3$ doped with 
non-magnetic impurities such as Zn, Mg or Si.
The main conclusions that we are able to draw are the following:

1- Whereas the investigation of the static susceptibility of doped 
CuGeO$_3$ enables us to define a universal behavior, in the proper reduced 
units, for the appearance of the paramagnetic, spin-Peierls and 
antiferromagnetic phases in the Temperature-Concentration diagram, at least 
two different regimes exist as concern the ESR properties, depending 
whether the doping atom is magnetic or not.

2- In the case of non-magnetic doping, the resulting Cu$^{2+}$ moments 
responsible for the Curie tail observed below the spin-Peierls temperature 
$T_{SP}$ are in strong coupling regime with the spin-Peierls excited 
triplet, resulting in an unshifted ESR line and a $c$-direction easy axis 
in the AFMR analysis.

3- In presence of doping with $S=1$ Ni$^{2+}$, a weak coupling is found 
from the analysis of the measured large shift of the Cu$^{2+}$ ESR, both in 
the paramagnetic and the spin-Peierls regimes.

4- However, in order to account for the AFMR $a$-easy axis in the latter 
case, one has to assume that the Ni$^{2+}$ ions do participate to the AFMR 
mode. The origin of the anisotropy in that case is the single ion 
anisotropy of the Ni$^{2+}$ rather than the anisotropic exchange present 
for pure and (non-magnetic) doped CuGeO$_3$.

5- Although each of these different regimes can be consistently 
parameterized in a quantitative way, a general description of the dynamics 
of the transverse spin susceptibilities of doped spin-Peierls compounds is 
still an open question.
 
\section{Acknowledgments}
We thank A. K. Hassan for helpfull discussions and for the high field EPR of CuGeO$_3$:Zn single crystals. This work was partially supported by the "MR Science Research Program" from 
RIKEN. We would like to thank M. Tokunaga for his help in the AFMR 
measurement. The Institut d'Electronique Fondamentale and Laboratoire de 
Physico-Chimie de l'Etat Solide are Unit\'es Mixtes de Recherche CNRS: UMR 
8622 and UMR 8648, respectively.

\vspace{0.5cm}
{\small \sl


\begin{references}
\bibitem{Grenier99}
Among the twelve investigated samples (doped with 1.6, 3.9 and 10\% Zn, 1, 
3 and 5\% Mg, 0.2, 0.7, 0.85, 1.5, 6 and 8.5\% Si), the linewidth was found 
to be always the same as in pure CuGeO$_3$ except for the 0.7\% Si- and 3.9\% 
Zn-doped samples, where a temperature independent constant broadening (of 
$\simeq$ 200 Gauss along $a$ and $c$, $\simeq$ 400 G along $b$) was 
measured.
\bibitem{note1}
Note that the results from the 0.2\% Ni-doped sample and the pure one can 
be directly compared on the same $T$ scale since the spin-Peierls 
temperatures are very close to each other (see Table I).
\bibitem{note2}
In Kittel's paper, due to the particular form of the damping specific to 
ferrimagnetic materials, a special case of strong coupling in presence of 
infinite relaxation rate is obtained. This results in a shift which is 
independent from the coupling contrary to Eq. (12) of the present work. The 
latter is simply related to the "unbottleneck" situation.
\end{references}
\end{document}